\documentclass[a4paper,11pt,preprint]{elsarticle}

\pdfoutput=1 
\usepackage{amssymb}
\usepackage{amsmath}

\usepackage[T1]{fontenc} 
\usepackage[lmargin=2.65cm, rmargin=2.65cm, bmargin=3.5cm]{geometry}
\usepackage{caption}
\begin{document} 

\begin{frontmatter}
\title{On the low-energy limit of the QED N-photon amplitudes: part 2}

\author[a]{James P. Edwards\corref{c1}}
\ead{jedwards@ifm.umich.mx}

\author[b]{Adolfo Huet}
\ead{adolfo.huet@uaq.mx}

%\author[b]{Cesar Moctezuma Mata}
%\ead{cesarfismat@gmail.com}

\author[a]{Christian Schubert}
\ead{schubert@ifm.umich.mx}

\address[a]{Instituto de F\'isica y Matem\'aticas, Universidad Michoacana de San Nicol\'as de
Hidalgo, Edificio C-3, Ciudad Universitaria.  Francisco J. M\'ujica s/n. Col. Fel\'icitas del R\'io.  58040 Morelia, Michoac\'an, M\'exico.}

\address[b]{Facultad de Ingenier\'ia, Universidad Aut\'onoma de Quer\'etaro, Cerro de las Campanas s/n, Colonia Las Campanas, Centro Universitario, 76010, Quer\'etaro, Quer\'etaro, M\'exico. }

\cortext[c1]{Corresponding author}

\begin{keyword}
Photon amplitudes \sep Quantum electrodynamics \sep Euler-Heisenberg Lagrangian \sep tadpole diagram  \\
\end{keyword}

\begin{abstract}
In recent work, Gies and Karbstein have discovered that the two-loop Euler-Heisenberg Lagrangians for scalar and spinor QED have non-vanishing reducible contributions
in addition to the well-studied irreducible ones. This invalidates previous applications of those Lagrangians to the computation of the two-loop $N$-photon amplitudes
in the low energy limit. Here we compute the corrections to those amplitudes due to the reducible contributions.  
\end{abstract}

\end{frontmatter}

\tableofcontents

%------------------------------------------------------------------------     
%COLORS
\def\green{\color{green}}
\def\blue{\color{blue}}
\def\red{\color{red}}
\def\black{\color{black}}
% MATH SYMBOLS
%

\def\veps{\varepsilon}
\newcommand{\Zz}{\mathcal{Z}}
\newcommand{\Zzp}{\mathcal{Z}^{\prime}}
\newcommand{\detZ}{\textrm{det}^{-\frac{1}{2}}\left[\frac{\sin\Zz}{\mathcal{\Zz}}\right]}
\newcommand{\detZp}{\textrm{det}^{-\frac{1}{2}}\left[\frac{\sin \Zzp}{\Zzp}\right]}
\newcommand{\detZs}{\textrm{det}^{-\frac{1}{2}}\left[\frac{\tan\Zz}{\mathcal{\Zz}}\right]}
\newcommand{\detZps}{\textrm{det}^{-\frac{1}{2}}\left[\frac{\tan\Zzp}{\Zzp}\right]}
\newcommand{\pdetZ}{\textrm{det}^{-\frac{1}{2}}\left[\cos \Zz \right]}
\newcommand{\pdetZp}{\textrm{det}^{-\frac{1}{2}}\left[\cos(\Zzp)\right]}

\newcommand{\tZz}{\frac{\tan\Zz}{\Zz}}
\newcommand{\tZzp}{\frac{\tan\Zzp}{\Zzp}}
\newcommand{\link}{\Big\vert_k}

\newcommand{\xm}{x_{-}}
\newcommand{\xp}{x_{+}}
\newcommand{\yp}{y_{+}}
\newcommand{\ym}{y_{-}}

\newcommand{\delC}{\underset{\smile}{\Delta}}
\newcommand{\ddelC}{{^{\bullet}\!\delC}}
\newcommand{\delCd}{{\delC\!^{\bullet}}}
\newcommand{\ddelCd}{{^{\bullet}\!\delC\!^{\bullet}}}
\newcommand{\odelC}{{^{\circ}\!\delC}}
\newcommand{\delCo}{{\delC\!^{\circ}}}
\newcommand{\odelCo}{{^{\circ}\!\delC\!^{\circ}}}
\newcommand{\odelCd}{{^{\circ}\!\delC\!^{\bullet}}}
\newcommand{\ddelCo}{{^{\bullet}\!\delC\!^{\circ}}}

\newcommand{\gb}{{\mathcal{G}_{B}}}
\newcommand{\gbd}{{\dot{\mathcal{G}}_{B}}}
\newcommand{\gbdm}{\dot{\mathcal{G}}_{B \mu\nu}}
\newcommand{\gf}{{\mathcal{G}_{F}}}
\newcommand{\gfd}{{\dot{\mathcal{G}}_{F}}}
\newcommand{\gfdm}{\dot{\mathcal{G}}_{F \mu\nu}}

\newcommand{\Fd}{\widetilde{F}}
\newcommand{\e}{\textrm{e}}
\newcommand{\Llsc}{\mathcal{L}^{(1)}_{\textrm{scal}}}
\newcommand{\Llsp}{\mathcal{L}^{(1)}_{\textrm{spin}}}
\newcommand{\cp}{\chi_{+}}
\newcommand{\cm}{\chi_{-}}
\newcommand{\nCr}[2]{\begin{pmatrix}#1 \\ #2\end{pmatrix}}

\def\arglist{[\varepsilon_1^+,k_1;\ldots ;\pol_K^+,k_K;\pol_{K+1}^-,k_{K+1};\ldots ;\pol_N^-,k_N]}

\def\cZ{{\cal Z}}

\def\cosech{\rm cosech}
\def\sech{\rm sech}
\def\coth{\rm coth}
\def\tanh{\rm tanh}
\def\tan{\rm tan}
%fractions
\def\half{{1\over 2}}
\def\third{{1\over3}}
\def\fourth{{1\over4}}
\def\fifth{{1\over5}}
\def\sixth{{1\over6}}
\def\seventh{{1\over7}}
\def\eigth{{1\over8}}
\def\ninth{{1\over9}}
\def\tenth{{1\over10}}
\def\conj{{{\rm c.c.}}}
\def\bN{\mathop{\bf N}}
\def\R{{\rm I\!R}}
\def\Eins{{\mathchoice {\rm 1\mskip-4mu l} {\rm 1\mskip-4mu l}
{\rm 1\mskip-4.5mu l} {\rm 1\mskip-5mu l}}}
\def\Z{{\mathchoice {\hbox{$\sf\textstyle Z\kern-0.4em Z$}}
{\hbox{$\sf\textstyle Z\kern-0.4em Z$}}
{\hbox{$\sf\scriptstyle Z\kern-0.3em Z$}}
{\hbox{$\sf\scriptscriptstyle Z\kern-0.2em Z$}}}}
\def\abs#1{\left| #1\right|}
\def\com#1#2{
        \left[#1, #2\right]}
%\def\square{\kern1pt\vbox{\hrule height 1.2pt\hbox{\vrule width 1.2pt
%   \hskip 3pt\vbox{\vskip 6pt}\hskip 3pt\vrule width 0.6pt}
%   \hrule height 0.6pt}\kern1pt}
 %     \def\boxop{{\raise-.25ex\hbox{\square}}}
% \contract is a differential geometry contraction sign _|
\def\contract{\makebox[1.2em][c]{
        \mbox{\rule{.6em}{.01truein}\rule{.01truein}{.6em}}}}
\def\ltap{\ \raisebox{-.4ex}{\rlap{$\sim$}} \raisebox{.4ex}{$<$}\ }
\def\gtap{\ \raisebox{-.4ex}{\rlap{$\sim$}} \raisebox{.4ex}{$>$}\ }
\def\mn{{\mu\nu}}
\def\rs{{\rho\sigma}}
\newcommand{\Det}{{\rm Det}}
\def\Tr{{\rm Tr}\,}
\def\tr{{\rm tr}\,}
\def\sumij{\sum_{i<j}}
\def\e{\,{\rm e}}
%boldface vectors
\def\br{{\bf r}}
\def\bp{{\bf p}}
\def\bq{{\bf q}}
\def\bx{{\bf x}}
\def\by{{\bf y}}
\def\brhat{{\bf \hat r}}
\def\bv{{\bf v}}
\def\ba{{\bf a}}
\def\bE{{\bf E}}
\def\bB{{\bf B}}
\def\bA{{\bf A}}
\def\b0{{\bf 0}}
%derivatives
\def\pa{\partial}
\def\dA{\partial^2}
\def\ddx{{d\over dx}}
\def\ddt{{d\over dt}}
\def\der#1#2{{d #1\over d#2}}
\def\lie{\hbox{\it \$}} % fancy L for the Lie derivative
\def\partder#1#2{\frac{\partial #1}{\partial #2}}
\def\secder#1#2#3{{\partial^2 #1\over\partial #2 \partial #3}}
%
%equations
%\newcommand{\be}{\blue\begin{equation}}
%\newcommand{\ee}{\end{equation}\black\noindent}
%\newcommand{\bear}{{\blue\begin{eqnarray}}}
%\newcommand{\ear}{{\end{eqnarray}\black\noindent}}
%\newcommand{\benn}{\begin{enumerate}}
%\newcommand{\enn}{\end{enumerate}}
%\newcommand{\veject}{\vfill\eject}
%\newcommand{\ven}{\vfill\eject\noindent}
\def\be{\begin{equation}}
\def\ee{\end{equation}\noindent}
\def\bear{\begin{eqnarray}}
\def\ear{\end{eqnarray}\noindent}
\def\bec{\blue\begin{equation}}
\def\eec{\end{equation}\black\noindent}
\def\bearc{\blue\begin{eqnarray}}
\def\earc{\end{eqnarray}\black\noindent}
\def\benn{\begin{enumerate}}
\def\enn{\end{enumerate}}
\def\veject{\vfill\eject}
\def\ven{\vfill\eject\noindent}
%
%reference to equations
\def\eq#1{{eq. (\ref{#1})}}
\def\eqs#1#2{{eqs. (\ref{#1}) -- (\ref{#2})}}
%
%algebra
\def\inv#1{\frac{1}{#1}}
\def\sumninf{\sum_{n=0}^{\infty}}
%
%integrals
\def\totint{\int_{-\infty}^{\infty}}
\def\posint{\int_0^{\infty}}
\def\negint{\int_{-\infty}^0}
\def\pint{{\dps\int}{dp_i\over {(2\pi)}^d}}
\def\intdp3{\int\frac{d^3p}{(2\pi)^3}}
\def\intdp4{\int\frac{d^4p}{(2\pi)^4}}
%propagators
\def\scalprop#1{\frac{-i}{#1^2+m^2-i\epsilon}}
%
% PHYS SYMBOLS
\newcommand{\GeV}{\mbox{GeV}}
\def\FFdual{F\cdot\tilde F}
\def\bra#1{\langle #1 |}
\def\ket#1{| #1 \rangle}
\def\braket#1#2{\langle {#1} \mid {#2} \rangle}
\def\vev#1{\langle #1 \rangle}
\def\matel#1#2#3{\langle #1\mid #2\mid #3 \rangle}
\def\rightvac{\mid0\rangle}
\def\leftvac{\langle0\mid}
\def\ihbar{{i\over\hbar}}
\def\lagr{{\cal L}}
% spinor stuff
\def\sigmabar{{\bar\sigma}}
% dirac matrix stuff
\def\ge{\hbox{$\gamma_1$}}
\def\gz{\hbox{$\gamma_2$}}
\def\gd{\hbox{$\gamma_3$}}
\def\go{\hbox{$\gamma_1$}}
\def\gt{\hbox{$\gamma_2$}}
\def\gth{\hbox{$\gamma_3$}} 
\def\gf{\hbox{$\gamma_5\;$}}
\def\slash#1{#1\!\!\!\raise.15ex\hbox {/}}
\newcommand{\slD}{\,\raise.15ex\hbox{$/$}\kern-.27em\hbox{$\!\!\!D$}}
\newcommand{\slpartial}{\raise.15ex\hbox{$/$}\kern-.57em\hbox{$\partial$}}
\newcommand{\PP}{\cal P}
\newcommand{\G}{{\cal G}}
\newcommand{\nc}{\newcommand}
\nc{\spa}[3]{\left\langle#1\,#3\right\rangle}
\nc{\spb}[3]{\left[#1\,#3\right]}
\nc{\ksl}{\not{\hbox{\kern-2.3pt $k$}}}
\nc{\hf}{\textstyle{1\over2}}
\nc{\pol}{\varepsilon}
\nc{\tq}{{\tilde q}}
\nc{\esl}{\not{\hbox{\kern-2.3pt $\pol$}}}
\newcommand{\cL}{\cal L}
\newcommand{\D}{\cal D}
\newcommand{\Dhalf}{{D\over 2}}
\def\eps{\epsilon}
\def\epshalf{{\epsilon\over 2}}
\def\lag{( -\partial^2 + V)}
%worldline
\def\freeexp{{\rm e}^{-\int_0^Td\tau {1\over 4}\dot x^2}}
\def\kinb{{1\over 4}\dot x^2}
\def\kinf{{1\over 2}\psi\dot\psi}
\def\expk{{\rm exp}\biggl[\,\sum_{i<j=1}^4 G_{Bij}k_i\cdot k_j\biggr]}
\def\expp{{\rm exp}\biggl[\,\sum_{i<j=1}^4 G_{Bij}p_i\cdot p_j\biggr]}
\def\expshort{{\e}^{\half G_{Bij}k_i\cdot k_j}}
\def\expabb{{\e}^{(\cdot )}}
\def\epseps#1#2{\varepsilon_{#1}\cdot \varepsilon_{#2}}
\def\epsk#1#2{\varepsilon_{#1}\cdot k_{#2}}
\def\kk#1#2{k_{#1}\cdot k_{#2}}
\def\G#1#2{G_{B#1#2}}
\def\Gp#1#2{{\dot G_{B#1#2}}}
\def\GF#1#2{G_{F#1#2}}
\def\Dab{{(x_a-x_b)}}
\def\Dsq{{({(x_a-x_b)}^2)}}
\def\PITD{{(4\pi T)}^{-{D\over 2}}}
\def\4piTD{{(4\pi T)}^{-{D\over 2}}}
\def\4piT4{{(4\pi T)}^{-2}}
\def\TintmD{{\dps\int_{0}^{\infty}}{dT\over T}\,e^{-m^2T}
    {(4\pi T)}^{-{D\over 2}}}
\def\Tintm4{{\dps\int_{0}^{\infty}}{dT\over T}\,e^{-m^2T}
    {(4\pi T)}^{-2}}
\def\Tintm{{\dps\int_{0}^{\infty}}{dT\over T}\,e^{-m^2T}}
\def\Tint{{\dps\int_{0}^{\infty}}{dT\over T}}
\def\np{n_{+}}
\def\nm{n_{-}}
\def\Np{N_{+}}
\def\Nm{N_{-}}
\newcommand{\slG}{{{\dot G}\!\!\!\! \raise.15ex\hbox {/}}}
\newcommand{\Gd}{{\dot G}}
\newcommand{\Gund}{{\underline{\dot G}}}
\newcommand{\Gdd}{{\ddot G}}
\def\GBd12{{\dot G}_{B12}}
\def\Dx{\dps\int{\cal D}x}
\def\Dy{\dps\int{\cal D}y}
\def\Dpsi{\dps\int{\cal D}\psi}
\def\dint#1{\int\!\!\!\!\!\int\limits_{\!\!#1}}
\def\ddtau{{d\over d\tau}}
\def\ie{\hbox{$\textstyle{\int_1}$}}
\def\iz{\hbox{$\textstyle{\int_2}$}}
\def\id{\hbox{$\textstyle{\int_3}$}}
\def\ldop{\hbox{$\lbrace\mskip -4.5mu\mid$}}
\def\rdop{\hbox{$\mid\mskip -4.3mu\rbrace$}}
%
%VARIOUS
\newcommand{\1}{{\'\i}}
\newcommand{\no}{\noindent}
\def\non{\nonumber}
\def\dps{\displaystyle}
\def\sy{\scriptscriptstyle}
\def\sy{\scriptscriptstyle}

\section{Introduction: the QED photon amplitudes}
\label{sec:intro}

Despite the remarkable progress that has been achieved in recent years in the calculation of 
on-shell amplitudes, particularly in the massless and/or SUSY cases (see, for example, 
\cite{berkos,bernreview,dixonreview,davyd,bedikopent,caglmi,pittau,weinzierl,fljeta,biguhe,bghs,elvhua}),
presently explicit calculations of loop amplitudes in gauge theory are, except for special helicity configurations,
still confined to a small number of particles. Even the prototypical QED one-loop $N$-photon amplitudes are currently
known only up to the six-point level \cite{karneu,cotopi,pasvel,denner}. 
For the massless case, there is also a vanishing theorem by Mahlon for the amplitudes
with $N\geq 6$ and all or all but one helicities equal \cite{mahlon} but less is known for the massive case. 

Things are very different if one wants only the low-energy limit of these amplitudes, i.e. for photon momenta
such that all kinematic invariants $k_i\cdot k_j$ are small compared to $m^2$.
In this limit the information on the photon amplitudes is fully contained in the effective Lagrangian ${\cal L}(F)$ for a background
field with a constant field strength tensor $F_{\mu\nu}$. The extraction of the low-energy amplitudes from the effective Lagrangian is straightforward in principle, 
and for the four-point case can be found in textbooks (for example see \cite{itzzubbook}).

At one loop, the QED effective Lagrangian for the constant field strength
case is just the well-known Euler-Heisenberg
Lagrangian \cite{eulhei} (see \cite{dunnerev} for a review), whose
weak field expansion is known in closed form.
In \cite{56} (called ``part one'' in the following) this expansion was used together
with the spinor helicity technique \cite{bkdgw,klesti,xuzhch} to arrive at a closed-form expression for the
one-loop $N$-photon amplitudes for any number of photons and any helicity distributions. 
This was also done in parallel for the scalar QED case, where the corresponding effective Lagrangian
is due to Weisskopf \cite{weisskopf}. 

In part one, this program was also carried to the two-loop level. 
Here none of the known representations of the two-loop effective Lagrangians in a constant field
\cite{ritusspin,ritusscal,ginzburg,ditreu-book,rss,frss,korsch}
is sufficiently explicit to obtain corresponding all - $N$ formulas
at the two-loop level. Nevertheless, the formulas given in \cite{ritusspin,ritusscal,ginzburg}
were good enough  to obtain the weak-field expansions of these
two-loop effective Lagrangians up to the order $F^{10}$, which 
allowed the explicit calculation of the two-loop $N$-photon amplitudes
up to the ten-point level with arbitrary helicities, in this low-energy limit. 

However, something special happens again for the ``all equal helicity'' amplitudes. 
For the effective action, those correspond to the special case of a self-dual field \cite{dufish,bardeen,selivanov,cangemi,chasie}
and for such a (constant) background it is possible to compute the effective action explicitly
even at the two loop level, for both scalar and spinor QED \cite{50,51}.
In \cite{51}, this fact was used to derive simple closed-form expressions for these ``all +'' amplitudes even at the two-loop level. 

A qualitative result of part 1 was the following ``double Furry theorem:''  while the
$N$-photon amplitudes corresponding to $K$ ($L$) helicity $+$ ($-$) photons with full energies restrict only the sum $K+L=N$ to be even,
in the low-energy limit both $K$ and $L$ have to be even, i.e. the amplitudes with $K$ or $L$ odd
vanish in this limit; thus the Euler-Heisenberg Lagrangian holds no information on them.  
This follows from a lack of non-vanishing invariants, and thus must hold at any loop order. 

\begin{figure}[hb!]
\begin{center}
 \includegraphics[width=0.35\textwidth]{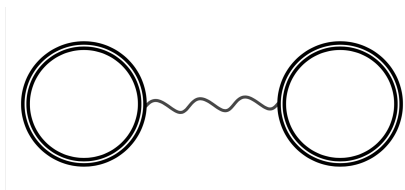}
\caption{{\bf One-particle reducible contribution to the two-loop EHL.}}
\label{fig-EHL1PR}
\end{center}
\end{figure}

A crucial point is that in these calculations to date it was assumed that the only diagram contributing to the EHL at the two-loop
level is the one particle irreducible (`1PI') one shown in Fig. \ref{fig-EHL1PI} (the double line denotes the full electron propagator in a constant field). At the same loop order, there is also the one-particle reducible (`1PR') diagram shown in Fig. \ref{fig-EHL1PR}. However, since the one-photon amplitude in a constant field formally vanishes on account of gauge invariance and momentum conservation, this 1PR diagram previously was generally discarded in the literature (see, e.g., \cite{ditreu-book,frgish-book}). 
However, Gies and Karbstein \cite{giekar} recently showed that this diagram actually gives a finite contribution,
if one takes into account the divergence of the connecting photon propagator in the zero-momentum limit. A careful analysis of that
limit led them to the following simple covariant formula that expresses this contribution to the two-loop Lagrangian in terms of derivatives of the one-loop Lagrangian:

\begin{figure}[t]
\begin{center}
 \includegraphics[width=0.16\textwidth]{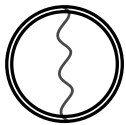}
\caption{{\bf One-particle irreducible contribution to the two-loop EHL.}}
\label{fig-EHL1PI}
\end{center}
\end{figure}

\bear
{\cal L}^{(2)\rm 1PR}
&=& \partder{{\cal L}^{(1)}}{F^\mn} \partder{{\cal L}^{(1)}}{F_\mn} \, .
\label{gieskarb}
\ear
This discovery has many consequences for constant-field QED, of which some have already been worked out,
namely the tadpole contributions to the one-loop propagators in scalar \cite{112} and spinor QED \cite{113}, as well as to the two-loop
photon vacuum polarization \cite{karbstein-vp}. In particular it renders incomplete all the results obtained in part 1
for the two-loop $N$-photon amplitudes, starting at the six-point level. The purpose of the present paper is to work out the changes to those results implied by the
non-vanishing of the reducible diagram.
 
In the next section, we will shortly summarize what was previously known about the (scalar and spinor) QED $N$-photon amplitudes in the low energy limit. 
To avoid undue repetition, here we will refer the reader to part 1 for some of the details. 
In section \ref{reducible} we give our results for the effect of the reducible diagram and tabulate updated coefficients taking these new contributions into account.

\section{The $N$-photon amplitudes in the low-energy limit: summary of known results}

Since in the abelian case the ordering of the legs does not
matter we assume that photons $1,\ldots, K$ carry the helicity `+' and the remaining $L$ photons the helicity `-'. Furthermore due to the double-Furry theorem mentioned above we can take both $K$ and $L$ to be even and we denote their sum by $K + L = N$.

\subsection{Low-energy photon amplitudes from Euler-Heisenberg Lagrangians: general procedure}

The extraction of the $N$-photon amplitudes from the effective action proceeds as follows: 
One chooses photon momenta $k_1,\ldots ,k_N$ and polarisation vectors
$\pol_1,\ldots ,\pol_N$, and defines for every leg the
field strength tensor

\bear
F_i^{\mu\nu} &\equiv & k_i^{\mu}\pol_i^{\nu} -  k_i^{\nu}\pol_i^{\mu} \, .
\ear
Then define the sum

\bear
F_{\rm tot} &\equiv & \sum_{i=1}^N F_i \,.
\label{defFtot}
\ear
The low-energy amplitude is obtained by inserting
$F_{\rm tot}$ into the effective Lagrangian, expanded to the appropriate order, 
and selecting the terms involving each $F_1,\ldots,F_N$ once:

\bear 
\Gamma
[k_1,\varepsilon_1;\ldots;k_N,\varepsilon_N]
&=&
{\cal L}\bigl(iF_{\rm tot}\bigr)\bigg\vert_{F_1\cdots F_N} \, .
\label{defEHamp}
\ear
For the four-photon case this is a standard textbook exercise \cite{itzzubbook}. To carry it out in the general $N$-photon case, it is convenient 
to use a helicity basis for the polarisations and apply the spinor helicity technique. An efficient method was developed in part 1: 
to obtain the amplitude $\Gamma^{(EH)}\arglist$
with $K$ positive-helivity photons and $L=N-K$ negative-helicity photons, the following steps should be taken:

\benn

\item
Replace $F$ by $iF$ in the effective Lagrangian ${\cal L}(F)$. 

\item
Rewrite the effective Lagrangian in terms of the invariants $a,\, b$, that are the invariants of the Maxwell field, defined by (as usual $\Fd_{\mu\nu} := \frac{1}{2}\epsilon_{\mu\nu\alpha\beta}F^{\alpha\beta}$)

\bear
a^2 &=&
\fourth \sqrt{\bigl(F_{\mu\nu}F^{\mu\nu}\bigr)^2
+\bigl(F_{\mu\nu}\tilde F^{\mu\nu}\bigr)^2}
+\fourth F_{\mu\nu}F^{\mu\nu} \, ,
\non\\
b^2 &=&
\fourth \sqrt{\bigl(F_{\mu\nu}F^{\mu\nu}\bigr)^2
+\bigl(F_{\mu\nu}\tilde F^{\mu\nu}\bigr)^2}
-\fourth F_{\mu\nu}F^{\mu\nu} \, ,
\non\\
\label{defab}
\ear
such that $a^2-b^2=B^2-E^2, (ab)^2 = ({\bf E}\cdot{\bf B})^2$.
The charge $e$ will often be set to unity in the following.

\item
Change variables from $a,b$ to $\chi_{\pm}$ via

\bear
a &=& \sqrt{\chi_+} + \sqrt{\chi_-} \, ,\non\\
b &=& -i(\sqrt{\chi_+}-\sqrt{\chi_-}) \, . \non\\
\label{ab}
\ear

\item
Expand the effective Lagrangian in powers of $\chi_+,\chi_-$. 

\item
Retain only the terms involving $\chi_+^{K\over 2}\chi_-^{L\over 2}$. This selects the contribution to the particle loop dressed by $K$ ($L$) low energy photons of helicity $+$ ($-$) from the constant background.

\item
In those, effect the replacement

\bear
\chi_+^{K\over 2}\to \chi_K^+ &\equiv & 
%(\chi_+)^{K\over 2}
%\Big\vert_{\rm all\,\, different}\non\\
%&=&
{({\frac{K}{2}})!
\over 2^{K\over 2}}
\Bigl\lbrace
[12]^2[34]^2\cdots [(K-1)K]^2 + {\rm \,\, all \,\, permutations} 
\Bigr\rbrace \, ,
\non\\
\chi_-^{L\over 2} \to \chi_{L}^- &\equiv & 
%(\chi_-)^{N-K\over 2}
%\Big\vert_{\rm all\,\, different}\non\\
%&=&
{({\frac{L}{2}})!
\over 2^{L\over 2}}
\Bigl\lbrace
\langle (K+1)(K+2)\rangle^2\langle (K+3)(K+4)\rangle^2\cdots
\langle (N-1)N\rangle^2 + {\rm \,\, all \,\, perm.} 
\Bigr\rbrace \, ,
\non\\
\label{defchiKL+-}
\ear
where $[ij]$ and $\langle ij\rangle$ are spinor products (our spinor helicity conventions follow \cite{dixonreview}). 

\enn

\subsection{One-loop $N$-photon amplitudes}

To summarise the existing results at one-loop level, we use the well-known integral representations of the effective Lagrangians due to Euler and Heisenberg \cite{eulhei} for spinor QED, and to Weisskopf \cite{weisskopf}
for scalar QED:

\bear
{\cal L}_{\rm spin}^{(1)}
&=&
-
{1\over 8\pi^2}
\int_0^{\infty}{dT\over T}
\,\e^{-m^2T}
\biggl\lbrack
{e^2ab\over \tanh(eaT)\tan(ebT)}
-{e^2\over 3}\bigl(a^2-b^2\bigr)
-{1\over T^2} 
\biggr\rbrack \, ,
\label{L1spin}
\\
{\cal L}_{\rm scal}^{(1)}
&=&
{1\over 16\pi^2}
\int_0^{\infty}{dT\over T}
\,\e^{-m^2T}
\biggl\lbrack
{e^2ab\over \sinh(eaT)\sin(ebT)}
+{e^2\over 6} (a^2-b^2) -{1\over T^2}
\biggr] \, .
\label{L1scal}
\ear
Here $T$ denotes the proper-time of the
loop fermion.
Using the Taylor series,

\bear
\frac{x}{{\rm tan}\, x} &=&
\sum_{n=0}^{\infty}(-1)^n{2^{2n}{\cal B}_{2n}\over (2n)!}x^{2n} \, ,
\label{taylorcot}\\
\frac{x}{\sin x}&=&
-\sum_{n=0}^{\infty}
(-1)^n{\bigl(2^{2n}-2\bigr){\cal B}_{2n}
\over (2n)!}x^{2n}
\label{taylorcosec}
\ear
(the ${\cal B}_{2n}$ are Bernoulli numbers) steps $1$-$4$ of the above procedure yield a power series expansion for the one-loop Euler-Heisenberg and Weisskopf Lagrangians
\begin{align}
	\mathcal{L}^{(1)}_{\textrm{spin}}(i F) &= -\frac{m^{4}}{8\pi^{2}} \sum_{N = 4}^{\infty}\left(\frac{2e}{m^{2}}\right)^{N}\sum_{K = 0}^{N}c^{(1)}_{\textrm{spin}}\left(\frac{K}{2}, \frac{N-K}{2}\right)\chi_{+}^{\frac{K}{2}}\chi_{-}^{\frac{N - K}{2}}\, , \label{L1SpinPower}\\
	\mathcal{L}^{(1)}_{\textrm{scal}}(i F) &= \,\,\frac{m^{4}}{16\pi^{2}} \sum_{N = 4}^{\infty}\left(\frac{2e}{m^{2}}\right)^{N}\sum_{K = 0}^{N}c^{(1)}_{\textrm{scal}}\left(\frac{K}{2}, \frac{N-K}{2}\right)\chi_{+}^{\frac{K}{2}}\chi_{-}^{\frac{N - K}{2}} \, ,
\label{L1ScalPower}
\end{align}
where both sums are over even numbers and the coefficients are given by

\bear
c_{\rm spin}^{(1)}\Bigl({K\over 2},\frac{N-K}{2}\Bigr)
&=&
(-1)^{N\over 2}(N-3)!
\sum_{k=0}^K\sum_{l=0}^{N-K}
(-1)^{N-K-l}
{{\cal B}_{k+l}{\cal B}_{N-k-l}
\over
k!l!(K-k)!(N-K-l)!}
\, ,
\label{c1spin}\\
c_{\rm scal}^{(1)}\Bigl({K\over 2},\frac{N-K}{2}\Bigr)
&=&
(-1)^{N\over 2}(N-3)!
\sum_{k=0}^K\sum_{l=0}^{N-K}
(-1)^{N-K-l}
{
\bigl(1-2^{1-k-l)}\bigr)
\bigl(1-2^{1-N+k+l}\bigr)
{\cal B}_{k+l}{\cal B}_{N-k-l}
\over
k!l!(K-k)!(N-K-l)!} \, .
\nonumber\\
\label{c1scal}
\ear
The remaining steps then yield the $N$-photon low energy scattering amplitudes (with $K$ positive helicities and $L$ negative helicities)

\bear
\Gamma_{\rm spin}^{(1)(EH)}\arglist
&=&
-{m^4\over 8\pi^2}
\Bigl({2e\over m^2}\Bigr)^N
\,c_{\rm spin}^{(1)}\Bigl({K\over 2},{N-K\over 2}\Bigr)
\chi_K^+\chi_{N-K}^- \, ,
\nonumber\\
\label{res1lspin}\\
\Gamma_{\rm scal}^{(1)(EH)}
\arglist
&=&
{m^4\over 16\pi^2}
\Bigl({2e\over m^2}\Bigr)^N
\,c_{\rm scal}^{(1)}\Bigl({K\over 2},{N-K\over 2}\Bigr)
\chi_K^+\chi_{N-K}^-\nonumber\\
\label{res1lscal}
\ear
(for $N\geq 4$). 
We note that the coeffcients $c_{\rm spin,scal}^{(1)}\Bigl({K\over 2},\frac{L}{2}\Bigr)$ are symmetric in their arguments, as is required by the CP invariance
of the QED photon amplitudes.  

\subsection{Two-loop $N$-photon amplitudes} 

At the two-loop level, in \cite{56} the integral representations given in \cite{ritusspin,ritusscal} were used 
to compute the weak-field expansion of the irreducible contributions up to the order $(F^{10})$. This yielded the low-energy photon amplitudes up
to ten-point order in the form 

\bear
\Gamma_{\rm spin}^{(2)(EH)}
[\varepsilon_1^+;\ldots \pol_K^+;\pol_{K+1}^-;\ldots ;\pol_N^-]
&=&
-{\alpha \pi m^4\over 8\pi^2}
\Bigl({2e\over m^2}\Bigr)^N
\,c_{\rm spin}^{(2)}\Bigl({K\over 2},{N-K\over 2}\Bigr)
\chi_K^+\chi_{N-K}^- \, ,
\label{res2lspin}
\\
\Gamma_{\rm scal}^{(2)(EH)}
[\varepsilon_1^+;\ldots \pol_K^+;\pol_{K+1}^-;\ldots ;\pol_N^-]
&=&\hphantom{-}
{\alpha\pi m^4\over 16\pi^2}
\Bigl({2e\over m^2}\Bigr)^N
\,c_{\rm scal}^{(2)}\Bigl({K\over 2},{N-K\over 2}\Bigr)
\chi_K^+\chi_{N-K}^- \, ,
\label{res2lscal}
\ear
with coefficients $c_{\rm spin}^{(2)}$ and $c_{\rm scal}^{(2)}$ given in Table 1 of \cite{56} and $\alpha = \frac{e^{2}}{4\pi}$ the usual fine structure constant. For the ``all +'' helicity case the following closed-form expressions can be obtained \cite{51}:

\bear
c^{(2)}_{\rm spin}\Bigl(n,0\Bigr) &=&
{1\over (2\pi)^2}\biggl\lbrace
\frac{2n-3}{2n-2}\,{\cal B}_{2n-2}
+3\sum_{k=1}^{n-1}
{{\cal B}_{2k}\over 2k}
{{\cal B}_{2n-2k}\over (2n-2k)}
\biggr\rbrace \, ,
\nonumber\\
c^{(2)}_{\rm scal}\Bigl(n,0\Bigr)&=&
{1\over (2\pi)^2}\biggl\lbrace
\frac{2n-3}{2n-2}\,{\cal B}_{2n-2}
+\frac{3}{2}\sum_{k=1}^{n-1}
{{\cal B}_{2k}\over 2k}
{{\cal B}_{2n-2k}\over (2n-2k)}
\biggr\rbrace \, ,
\non\\
\label{allplus}
\ear
where $n=N/2$. 
In the following we update these results by including the one particle reducible contribution to the two-loop Euler-Heisenberg and Weisskopf Lagrangians.

\section{Two-loop: one-particle reducible contributions}
\label{reducible}
Here our aim is to express the reducible contribution in terms of kinematic invariants and to relate the two-loop, reducible contributions to the $N$-photon amplitudes to the one-loop coefficients reviewed above. 

\subsection{The reducible Lagrangian}
To write the reducible contribution in a manifestly Lorentz invariant way we use the standard invariants $a$ and $b$ defined in (\ref{defab}), inverting these expressions to write the field strength tensor and its dual as
\begin{equation}
	F^{2} = 2(a^{2} - b^{2})\,, \qquad (F \Fd)^2 = (4 a b)^2 \, .
\end{equation}
Moreover, one notes that $2(a^{2} + b^{2}) = \sqrt{(F^{2})^{2} + (F \Fd)^{2}}$. With the further results

\begin{align}
	\frac{\partial a^{2}}{\partial F_{\mu\nu}} &= \frac{1}{2}\left(F^{\mu\nu} + \frac{1}{2} \frac{F^{2}F^{\mu\nu} + F\Fd\, \Fd^{\mu\nu}}{a^{2} + b^{2}}\right) \, , \\
	\frac{\partial b^{2}}{\partial F_{\mu\nu}} &= \frac{1}{2}\left(-F^{\mu\nu} + \frac{1}{2} \frac{F^{2}F^{\mu\nu} + F\Fd\, \Fd^{\mu\nu}}{a^{2} + b^{2}}\right) 
\end{align}
we can express the covariant formula for the reducible contribution as (the cross terms vanish)

\begin{align}
	16\frac{\partial {\mathcal{L}}^{(1)}}{\partial F_{\mu\nu}}\frac{\partial {\mathcal{L}}^{(1)}}{\partial F^{\mu\nu}} &= \frac{1}{a^{2}}\left[F^{2} + \frac{(F^{2})^{2} + (F \Fd)^{2}}{a^{2} + b^{2}} + \frac{1}{4}\frac{F^{2}\left( (F^{2})^{2} + (F\Fd)^{2}\right)}{(a^{2} + b^{2})^{2}}\right] \left(\frac{\partial {\mathcal{L}}^{(1)}}{\partial a}\right)^{2}, \nonumber \\
	&+ \frac{1}{b^{2}}\left[F^{2} - \frac{(F^{2})^{2} + (F \Fd)^{2}}{a^{2} + b^{2}} + \frac{1}{4}\frac{F^{2}\left( (F^{2})^{2} + (F\Fd)^{2}\right)}{(a^{2} + b^{2})^{2}}\right]\left(\frac{\partial {\mathcal{L}}^{(1)}}{\partial b}\right)^{2} .
\end{align}
Rewriting the field strengths in terms of $a$ and $b$ leads to a re-writing of (\ref{gieskarb}) in the form

\begin{equation}
	\frac{\partial {\mathcal{L}}^{(1)}}{\partial F_{\mu\nu}}\frac{\partial {\mathcal{L}}^{(1)}}{\partial F^{\mu\nu}} = \frac{1}{2}\left[ \left(\frac{\partial {\mathcal{L}}^{(1)}}{\partial a}\right)^{2} - \left(\frac{\partial {\mathcal{L}}^{(1)}}{\partial b}\right)^{2} \right].
	\label{twoloopred}
\end{equation}
Note that this result is valid for either the spinor or scalar Lagrangian. To proceed we use the explicit formula for $\Llsc$ and $\Llsp$ in (\ref{L1scal}) and (\ref{L1spin}).
For spinor QED the derivatives of $\Llsp$ with respect to $a$ and $b$ are

\begin{align}
	\frac{\partial \Llsp}{\partial a} &= -\frac{e^{2}}{8 \pi^{2}}\int_{0}^{\infty} \frac{dT}{T}\, \e^{-m^{2} T} \left[\frac{b}{\tanh(e a T) \tan(e b T)} - \frac{2 e a b T}{\sinh(2 e a T)\tanh(e a T) \tan(e b T)  } - \frac{2a}{3}\right] \, ,\\
	\frac{\partial \Llsp}{\partial b} &= -\frac{e^{2}}{8 \pi^{2}}\int_{0}^{\infty} \frac{dT}{T}\, \e^{-m^{2} T} \left[\frac{a}{\tanh(e a T) \tan(e b T)} - \frac{2 e a b T}{\sin(2 e b T)\tanh(e a T) \tan(e b T)  } + \frac{2b}{3}\right] \, ,
\end{align}
so that squaring these and taking their difference yields

\begin{align}
&	\hspace{-16em}	\frac{\partial \Llsp}{\partial F_{\mu\nu}}\frac{\partial \Llsp}{\partial F^{\mu\nu}} = \nonumber \\
\hspace{-3em}	\frac{e^{4}}{128\pi^{4}}\int_{0}^{\infty} \frac{dT}{T} \int_{0}^{\infty}\frac{dS}{S} \,\e^{-m^{2}(T + S)} \Bigg\{ &  \frac{b^{2}\left(1 - \frac{2 e a T}{\sinh(2 e a T)}\right)\left(1 - \frac{2 e a S}{\sinh(2 e a S)}\right)- a^{2}\left(1 - \frac{2 e b T}{\sin(2 e b T)}\right)\left(1 - \frac{2 e b S}{\sin(2 e b S)}\right)}{\tanh(e a T)\tanh(e a S)\tan(e b T) \tan(e b S)}  \nonumber \\
-& \frac{4 a b}{3} \,
\biggl\lbrack
\frac{1 -  \frac{ e a T}{\sinh(2 e a T)} - \frac{ e b T}{\sin(2 e b T)} }{\tanh(e a T)\tan(e b T)}	
+
(T\to S )
\biggr\rbrack
+4 \frac{a^{2} - b^{2}}{9}\Bigg\}.
\end{align}
This formula can be expanded or evaluated numerically. The same procedure can be used for scalar QED, leading to very similar formulas. 

However, rather than finding explicit formulae for the two-loop coefficients from these proper time representations, in the next subsection we instead express them in terms of the one-loop coefficients.

\subsection{Reducible coefficients in terms of one-loop coefficients}
To relate the two-loop coefficients to their one-loop counterparts we make use of (\ref{res1lspin}) and (\ref{res1lscal}) that express the one-loop Lagrangian in terms of $\chi_{\pm}$. As above, we want to change variables in the covariant formula (\ref{gieskarb}), this time to $\chi_{\pm}$ which will allow us to manipulate the series representations of the one-loop Lagrangians directly. So we note that with $8\chi_{\pm} = F^{2} \pm i F \Fd$ derivatives with respect to the field strength can be converted to derivatives with respect to $\chi_{\pm}$,

\begin{equation}
\frac{\partial}{\partial F_{\mu\nu}} = \frac{1}{4}(F^{\mu\nu} + i \Fd^{\mu\nu})\frac{\partial }{\partial \cp} + \frac{1}{4}(F^{\mu\nu} - i \Fd^{\mu\nu})\frac{\partial}{\partial \cm}.
\end{equation}
From here it is straightforward to derive

\begin{equation}
	\frac{\partial }{\partial F_{\mu\nu}}\left(\cp^{\frac{k}{2}}\cm^{\frac{N - k}{2}}\right) = \frac{k}{8}(F^{\mu\nu} + i \Fd^{\mu\nu})\cp^{\frac{k-2}{2}}\cm^{\frac{N - k}{2}} + \frac{N-k}{8}(F^{\mu\nu} - i \Fd^{\mu\nu})\cp^{\frac{k}{2}}\cm^{\frac{N - k-2}{2}} \, .
\end{equation}
Squaring this and rewriting $F^{2}$ and $F \Fd$ in terms of $\chi_{\pm}$ provides (again the cross terms vanish)

\begin{eqnarray}
\frac{\partial }{\partial F_{\mu\nu}}\left(\cp^{\frac{k_1}{2}}\cm^{\frac{N_1 - k_1}{2}}\right)
\frac{\partial }{\partial F^{\mu\nu}}\left(\cp^{\frac{k_2}{2}}\cm^{\frac{N_2 - k_2}{2}}\right)
&=& \nonumber\\ 	
&& \hspace{-200pt}	\frac{1}{4}\left[k_{1}k_{2}\, \cp^{\frac{k_{1} + k_{2} - 2}{2}}\cm^{\frac{N_{1} + N_{2} -(k_{1} + k_{2})}{2}} + (N_{1} - k_{1})(N_{2} - k_{2})\, \cp^{\frac{k_{1} + k_{2}}{2}}\cm^{\frac{N_{1} + N_{2} - (k_{1} + k_{2}) - 2}{2}} \right].
	\label{dFC}
\end{eqnarray}
Beginning with spinor QED, (\ref{dFC}) enters the summand of the reducible contribution by applying it to the power series representation of the one-loop Lagrangian in (\ref{L1SpinPower}) (recall the sums over $N_{1}$, $N_{2}$ and $k_{1}$, $k_{2}$ are over even integers only):

\begin{align}
	\hspace{-1em}\frac{\partial \Llsp}{\partial (iF_{\mu\nu})}\frac{\partial \Llsp}{\partial (iF^{\mu\nu})} =&
- \frac{m^{8}}{256 \pi^{4}}\sum_{N_{1} = 4}^{\infty} \sum_{N_{2} = 4}^{\infty}\left(\frac{2 e}{m^{2}}\right)^{N_{1} + N_{2}} \sum_{k_{1} = 0}^{N_{1}}\sum_{k_{2} = 0}^{N_{2}} c^{(1)}_{\textrm{spin}}\left(\frac{k_{1}}{2}, \frac{N_{1} - k_{1}}{2}\right) c^{(1)}_{\textrm{spin}}\left(\frac{k_{2}}{2}, \frac{N_{2} - k_{2}}{2}\right) \nonumber \\
 & \times
  \left[k_{1} k_{2} \cp^{\frac{k_{1} + k_{2} - 2}{2}}\cm^{\frac{N_{1} + N_{2} - (k_{1} + k_{2})}{2}} + (N_{1} - k_{1})(N_{2} - k_{2})\cp^{\frac{k_{1} + k_{2}}{2}} \cm^{\frac{N_{1} + N_{2} - (k_{1} + k_{2}) - 2}{2}} \right].
 \label{redspin2}
\end{align}
Here the one-loop coefficients for spinor QED are given in (\ref{c1spin}). Note that by changing variables $k_{i} \rightarrow N_{i} - k_{i}$ for $i = 1, 2$ and 
using the symmetry of those coefficients with respect to interchange of their arguments, we can write this more concisely as

\begin{align}
	\hspace{-1em}\frac{\partial \Llsp}{\partial (i F_{\mu\nu})}\frac{\partial \Llsp}{\partial (iF^{\mu\nu})} =&
- \frac{m^{8}}{256 \pi^{4}}\sum_{N_{1} = 4}^{\infty} \sum_{N_{2} = 4}^{\infty}\left(\frac{2 e}{m^{2}}\right)^{N_{1} + N_{2}} \sum_{k_{1} = 0}^{N_{1}}\sum_{k_{2} = 0}^{N_{2}} 
 k_1 c^{(1)}_{\textrm{spin}}\left(\frac{k_{1}}{2}, \frac{N_{1} - k_{1}}{2}\right) k_2 c^{(1)}_{\textrm{spin}}\left(\frac{k_{2}}{2}, \frac{N_{2} - k_{2}}{2}\right) \nonumber \\
 \times & \left[ \cp^{\frac{k_{1} + k_{2} - 2}{2}}\cm^{\frac{N_{1} + N_{2} - (k_{1} + k_{2})}{2}} +
 \Bigl( \cp \leftrightarrow \cm\Bigr) \right].
 \label{redspin2sym}
\end{align}
This form has the further advantage of explicitly displaying the CP invariance of these reducible two-loop contributions. 

For the two-loop coefficient appropriate to the scattering of $K$ ($L$) low energy helicity plus (minus) photons we look for the terms proportional to $\cp^{\frac{K}{2}}\cm^{\frac{L}{2}}$. For the first term in \eqref{redspin2sym} 
we can fix $N_{2} = K + L + 2 - N_{1}$ as long as $N_{1} \leqslant K + L - 2$, and set $k_{2} = K + 2 - k_{1}$ subject to $K+2 \geqslant k_{1}$ and $k_{1} \geqslant N_{1} - L$. Then with the same normalisation as in (\ref{res2lspin}) we find a contribution to the two loop coefficient, $c^{(2)}_{\textrm{spin}}$, equal to 

\begin{eqnarray}
{\tilde c^{(2,\textrm{red})} }_{\textrm{spin}}\left(\frac{K}{2} , \frac{L}{2}\right)&=&\frac{1}{2 \pi^{2}}\sum_{N_{1} = 4}^{K+L-2} \sum_{k_{1} = \textrm{max}(N_{1} - L,\, 0)}^{\textrm{min}(N_{1}, K+2)}  c^{(1)}_{\textrm{spin}}\left(\frac{k_{1}}{2}, \frac{N_{1} - k_{1}}{2}\right) c^{(1)}_{\textrm{spin}}\left(\frac{K+2 - k_{1}}{2}, \frac{L - N_{1} + k_{1}}{2}\right) 
\nonumber\\
&&\times k_{1}(K+2-k_{1}),
\label{tildec}
\end{eqnarray}
recalling that for the $k_{1}$ summation one takes the first even integer satisfying the lower condition. The full reducible coefficient is obtained by the addition of the second term
in \eqref{redspin2sym}, amounting to a symmetrization in $K$ and $L$:

\bear
c^{(2,\textrm{red}) }_{\textrm{spin}}\left(\frac{K}{2} , \frac{L}{2}\right) = \tilde c^{(2,\textrm{red}) }_{\textrm{spin}}\left(\frac{K}{2} , \frac{L}{2}\right) 
+ \tilde c^{(2,\textrm{red}) }_{\textrm{spin}}\left(\frac{L}{2} , \frac{K}{2}\right)  \, .
\label{decomp}
\ear
Thus knowledge of the one-loop coefficients is sufficient to determine the reducible contribution to the two-loop coefficients.
\begin{table}
    \centering
 \begin{tabular}{c c c c }

 \hline  \\ 
 $ \left( \frac{K}{2}, \frac{N-K}{2} \right)$ & $c^{(2,{\rm irr})}_{\rm spin}$  & $c^{(2,{\rm red})}_{\rm spin}$& $c^{(2)}_{\rm spin}$\\ [1.5ex] 
 \hline\hline
 $\left( 5, 0 \right)$ &  $\frac{317}{40320\pi^2}$ & $\frac{467}{4233600\pi^2}$ & $\frac{4219}{529200\pi^2}$\\ [1.5ex]
 
 $ \left( 4, 1 \right)$  &  $\frac{-8707}{1814400\pi^2}$ & $\frac{-12241}{38102400\pi^2}$ & $\frac{-12193}{2381400\pi^2}$\\ [1.5ex]
 
 $ \left( 3, 2 \right)$  &  $\frac{-3190547}{8164800\pi^2}$ & $\frac{14837}{2721600\pi^2}$ & $\frac{-786509}{2041200\pi^2}$\\ [1.5ex]
 
 $ \left( 4, 0 \right)$ &  $\frac{2221}{403200\pi^2}$ & $\frac{1}{10080\pi^2}$ & $\frac{323}{57600\pi^2}$\\[1.5ex] 
 
 $\left( 3, 1 \right)$  &  $\frac{-151379}{6350400\pi^2}$ & $\frac{1}{22680\pi^2}$ & $\frac{-151099}{6350400\pi^2}$\\[1.5ex] 
 
 $\left( 2, 2 \right)$  &  $\frac{-37763}{282240\pi^2}$ & $\frac{703}{226800\pi^2}$ & $\frac{-1659967}{12700800\pi^2}$\\[1.5ex]
 
 $ \left( 3, 0 \right)$  &  $\frac{7}{960\pi^2}$ & $\frac{1}{7200\pi^2}$ & $\frac{107}{14400\pi^2}$\\[1.5ex]
 
 $ \left( 2, 1 \right)$  &  $\frac{-5821}{129600\pi^2}$ & $\frac{11}{12960\pi^2}$ & $\frac{-5711}{129600\pi^2}$\\[1.5ex]

 $ \left( 2, 0 \right)$  &  $\frac{5}{192\pi^2}$ & $0$ & $\frac{5}{192\pi^2}$\\[1.5ex]
 
 $ \left( 1, 1 \right)$  &  $\frac{-391}{2592\pi^2}$ & $0$ & $\frac{-391}{2592\pi^2}$\\ [1ex] 
 \hline

\end{tabular}  \caption{{\bf Coefficients for the spinor two-loop EHL.}}\label{SPINTABLE}
\end{table}

For scalar QED, the process is the same, and leads to the same formulas \eqref{tildec}, \eqref{decomp}, where the one-loop coefficients are now given by  (\ref{c1scal}).
However, care must be taken in the overall normalisation of the amplitudes. For the irreducible contribution we had a single scalar/spinor loop, and we chose to leave the
corresponding factor of $-2$ accounting for the difference in statistics and degrees of freedom as a global factor, rather than absorbing it into the coefficients $c^{(2,\textrm{irr})}$
(see \eqref{res2lspin}, \eqref{res2lscal}). 
For the reducible contribution we have two such factors of $-2$, and one of them must now be absorbed into the coefficients for consistency. 
Thus to obtain the scalar QED equivalent of eq. \eqref{tildec} we must, in addition to changing the one-loop coefficients, also replace the global prefactor $\frac{1}{2 \pi^{2}}$
by $-\frac{1}{4 \pi^{2}}$.

Finally, tables (\ref{SPINTABLE}) and (\ref{SCALTABLE}) show explicit numerical values for the two-loop coefficients up to order $F^{10}$. These correct the corresponding tables presented in \cite{56} that included only the irreducible contributions, making explicit the contribution of each of the two diagrams (irreducible and reducible) and the new 
total values for the coefficients. 

\begin{table}
    \centering %%SCALAR TABLE
 \begin{tabular}{c c c c } 
 \hline  \\ 
 $ \left( \frac{K}{2}, \frac{N-K}{2} \right)$  & $c^{(2,{\rm irr})}_{\rm scal} $ & $c^{(2,{\rm red})}_{\rm scal} $& $ c^{(2)}_{\rm scal}$\\ [1.5ex] 
 \hline\hline
 $\left( 5, 0 \right)$  &  $\frac{611}{80640\pi^2}$ & $\frac{-467}{8467200\pi^2}$ & $\frac{7961}{1058400\pi^2}$\\ [1.5ex]
 
 $ \left( 4, 1 \right)$  &  $\frac{349609}{3628800\pi^2}$ & $\frac{-449}{762048\pi^2}$ & $\frac{7296889}{76204800\pi^2}$\\ [1.5ex]
 
 $ \left( 3, 2 \right)$ &   $\frac{688637}{2332800\pi^2}$ & $\frac{-4507}{2721600\pi^2}$ & $\frac{4793417}{16329600\pi^2}$\\ [1.5ex]
 
 $ \left( 4, 0 \right)$  &  $\frac{67}{12800\pi^2}$ & $\frac{-1}{20160\pi^2}$ & $\frac{4181}{806400\pi^2}$\\ [1.5ex]
 
 $\left( 3, 1 \right)$  &  $\frac{273619}{6350400\pi^2}$ & $\frac{-1}{2835\pi^2}$ & $\frac{271379}{6350400\pi^2}$\\ [1.5ex]
 
 $\left( 2, 2 \right)$  &  $\frac{2055163}{25401600\pi^2}$ & $\frac{-143}{226800\pi^2}$ & $\frac{2039147}{25401600\pi^2}$\\ [1.5ex]
 
 $ \left( 3, 0 \right)$  &  $\frac{13}{1920\pi^2}$ & $\frac{-1}{14400\pi^2}$ & $\frac{193}{28800\pi^2}$\\ [1.5ex]

 $ \left( 2, 1 \right)$  &  $\frac{8563}{259200\pi^2}$ & $\frac{-1}{3240\pi^2}$ & $\frac{8483}{259200\pi^2}$\\ [1.5ex]
 
 $ \left( 2, 0 \right)$  &  $\frac{3}{128\pi^2}$ & $0$ & $\frac{3}{128\pi^2}$\\ [1.5ex]
 
 $ \left( 1, 1 \right)$ &  $\frac{307}{5184\pi^2}$ & $0$ & $\frac{307}{5184\pi^2}$\\ [1ex] 
 \hline
\end{tabular}  
\caption{{\bf Coefficients for the scalar two-loop EHL.}}
\label{SCALTABLE}
\end{table}

Let us note two properties of the new contributions from the reducible diagrams. First, they start contributing only from the six-photon level.
This is because the renormalized one-loop Lagrangians start only at the four-photon level, thus their square only at the eight-photon level,
and taking two derivatives lowers the starting point to the six-photon level. 
Second, the ``all +'' coefficients $c^{2,\textrm{red}}(N/2,0)$ come in a fixed ratio of $-2$ between the spinor and scalar cases. 
The reason is that those involve only the coefficients of the one-loop ``all +'' Lagrangians, and those come from the self-dual scalar and spinor Lagrangians 
that coincide at one loop (after renormalization and up to the global normalization, but the latter difference has been eliminated by our convention for the coefficients). 
This is due to the fact that  the Dirac operator in a self-dual field has a supersymmetry \cite{dufish,thooft,dadda,brolee}. 

In this special case that all photons have equal helicities our formulas above simplify considerably.  
Setting $L=0$, we find that only the first term in the decomposition \eqref{decomp} contributes in this case, and that \eqref{tildec} simplifies to 

%\begin{align}
%	\hspace{-2em}\sum_{k_{1} = N_{1} - 2}^{N_{1}}&(N_{1} - k_{1})(K_{1} + 2 - N_{1})c^{(1)}\left(\frac{k_{1}}{2}, \frac{N_{1} - k_{1}}{2}\right)c^{(1)}\left(\frac{K - k_{1}}{2}, \frac{k_{1} + 2 - N_{1}}{2}\right) \nonumber \\
%	=~& c^{(1)}\left(\frac{N_{1} - 1}{2}, \frac{1}{2}\right)c^{(1)}\left(\frac{K + 1 - N_{1}}{2}, \frac{1}{2}\right),
%\end{align} 
%where the coefficients are those for spinor or scalar QED as appropriate, provided this time that $N_{1}$ be odd. 

\begin{align}
\tilde c^{(2,\textrm{red}) }_{\textrm{spin}}\left(\frac{K}{2} , 0\right)	
=  \frac{1}{2\pi^{2}} \sum_{N_{1} = 4}^{K-2} \bigg[N_{1} &\left(K + 2 - N_{1}\right) c^{(1)}_{\textrm{spin}}\left(\frac{K+2 - N_{1}}{2}, 0\right)c^{(1)}_{\textrm{spin}}\left(\frac{N_{1}}{2}, 0\right) \bigg] \, . \nonumber\\
\end{align}
Now using the fact that according to our normalisations in (\ref{L1SpinPower}) and (\ref{L1ScalPower}) the all plus / all minus coefficients for scalar and spinor QED coincide as

\begin{equation}
	c^{(1)}_{\textrm{spin}}\left(n, 0\right) = c^{(1)}_{\textrm{scal}}\left(n, 0\right) = \frac{(-1)^{n+1}{\cal B}_{2n}}{2n(2n-2)} 
\end{equation}
with $n = \frac{N}{2}$ as in (\ref{allplus}), we can give the above results in a form similar to \eqref{allplus}:

\begin{align}
	c^{(2, \textrm{~red})}_{\textrm{spin}}\left(n, 0\right) &= \frac{(-1)^{n+1}}{2\pi^{2}}\sum_{m=2}^{n-1} \frac{{\cal B}_{2m}{\cal B}_{2(n-m+1)}}{(2m - 2)(2(n-m+1)-2)}\, ,\\
	c^{(2, \textrm{~red})}_{\textrm{scal}}\left(n, 0\right) &= -\frac{c^{(2,\textrm{~red})}_{\textrm{spin}}\left(n, 0\right)}{2} \, .
\end{align}
These results are in agreement with the entries reported in the tables above for $L = 0$. Analogous results can be found for coefficients with two minus photons by setting $L = 2$ and repeating the above analysis. 

\section{Summary and outlook}
We have worked out here the changes to the results of \cite{56} on the two-loop $N$-photon amplitudes in the low energy limit, necessary to take into account the 
recently discovered non-vanishing of the reducible contribution to the two-loop QED effective Lagrangian in a constant field. Contrary to the irreducible contributions,
the ``new'' reducible ones can be given in closed form for all helicity assignments, written in terms of the known one-loop coefficients. 

The formulas for the reducible contributions are written in a form advantageous for carrying out an asymptotic analysis. An analysis of special cases like "all-plus-helicities" or equal numbers of plus and minus helicities is also feasible. Consequences for the imaginary part of the effective Lagrangian and Schwinger 
pair creation will be given elsewhere. Furthermore, these reducible contributions will be present at all higher loop orders too, so that future work on such calculations must take them into consideration.

\subsection*{Acknowledgements}

C.S. thanks CONACyT for financial support through grant Ciencias Basicas 2014 No. 242461. J.P.E receives funding from PRODEP. Adolfo Huet acknowledges support from CONACyT.



\begin{thebibliography}{99}


\bibitem{berkos}
Z. Bern and D. A. Kosower,
Phys. Rev. Lett. {\bf 66} (1991) 1669;
Nucl. Phys. {\bf B 379} (1992)
451.

\bibitem{bernreview}
Z. Bern, L.J. Dixon, D.A. Kosower,
Ann. Rev. Nucl. Part. Sci. {\bf 46} (1996) 109,
hep-ph/9602280.

\bibitem{dixonreview}
L. Dixon,
TASI Lectures, Boulder TASI 95, 539, hep-ph/9601359.

\bibitem{davyd}
A.I. Davydychev, Phys. Lett. {\bf B 263} (1991) 107.

\bibitem{bedikopent}
Z.~Bern, L.~J.~Dixon and D.~A.~Kosower,
Nucl. Phys. {\bf B 412} (1994) 751,
hep-ph/9306240.

\bibitem{caglmi}
J.M. Campbell, E.W.N. Glover, D.J. Miller,
Nucl. Phys. {\bf B498} (1997) 397, hep-ph/9612413.

\bibitem{pittau}
R. Pittau, Comp. Phys. Comm. {\bf 104} (1997) 23,
hep-ph/9607309.

\bibitem{weinzierl}
S. Weinzierl, Phys. Lett. {\bf B 450} (1999) 234,
hep-ph/9811365.

\bibitem{fljeta}
J. Fleischer, F. Jegerlehner, and O.V. Tarasov,
Nucl. Phys. {\bf B 566} (2000) 423, hep-ph/9907327.

\bibitem{biguhe}
T.~Binoth, J.~P.~Guillet and G.~Heinrich,
Nucl. Phys. {\bf B 572} (2000) 361,
hep-ph/9911342.

\bibitem{bghs}
T.~Binoth, J.~P.~Guillet, G.~Heinrich and C.~Schubert,
Nucl. Phys. {\bf B 615} (2001) 385,
hep-ph/0106243.

\bibitem{elvhua}
H. Elvang and Y.-t. Huang,
%``Scattering Amplitudes'',
arXiv:1308.1697 [hep-th]. 

\bibitem{karneu}
R. Karplus and M. Neuman, Phys. Rev. {\bf 80} (1950) 380.

\bibitem{cotopi}
V. Costantini, D. De Tollis, G. Pistoni,
Nuov. Cim. {\bf 2A} (1971) 733.

\bibitem{pasvel}
G. Passarino, M. Veltman, Nucl. Phys. {\bf B 160} (1979) 151.

\bibitem{denner}
A. Denner, Forts. Phys. {\bf 41} (1993) 307.

\bibitem{mahlon} G. Mahlon,
Phys. Rev. D {\bf 49} (1994) 2197, hep-ph/9311213.

\bibitem{itzzubbook}
C. Itzykson, J. Zuber, {\it Quantum Field Theory},
McGraw-Hill 1985.

\bibitem{eulhei}
W. Heisenberg and H. Euler,
% ``Folgerungen aus derDiracschen Theorie des Positrons'', 
Z. Phys. {\bf 98} (1936) 714.

\bibitem{dunnerev}
G. V. Dunne, 
%``Heisenberg-Euler effective Lagrangians: Basics and extensions'', 
Ian Kogan Memorial Collection, {\sl From Fields to Strings: Circumnavigating Theoretical Physics}, 
M.A. Shifman et al (eds.) (2004), Vol. I, 445, arXiv:hep-th/0406216. 

\bibitem{56}
L. C. Martin, C. Schubert and V. M. Villanueva Sandoval, 
%ï¿œOn the low-energy limit of the QED N - photon amplitudesï¿œ, 
Nucl. Phys. B {\bf 668} (2003) 335, arXiv:hepth/0301022. 


\bibitem{bkdgw}
F.A. Berends, R. Kleiss, P. De Causmaecker, R. Gastmans, T.T. Wu,
Phys. Lett. {\bf B 103} (1981 ) 124.

\bibitem{klesti}
R. Kleiss, W.J. Stirling,
Nucl. Phys. {\bf B 262} (1985) 235.

\bibitem{xuzhch}
Z. Xu, D.-H. Zhang, L. Chang,
Nucl. Phys. {\bf B 291} (1987) 392.

\bibitem{weisskopf} V. Weisskopf,
Kong. Dans. Vid. Selsk.
Math-fys. Medd. XIV No. 6 (1936),
reprinted in {\it Quantum Electrodynamics}, J. Schwinger (Ed.) (Dover,
New York, 1958).



\bibitem{ritusspin}
V. I. Ritus, 
%``Lagrangian of an intense electromagnetic field and quantum electrodynamics at short distances'', 
Zh. Eksp. Teor.
Fiz {\bf 69} (1975) 1517 [Sov. Phys. JETP {\bf 42} (1975) 774].



\bibitem{ritusscal}
V. I. Ritus,
Zh. Eksp. Teor. Fiz {\bf 73} (1977) 807
[Sov. Phys. JETP {\bf 46} (1977) 423].

\bibitem{ginzburg}
V. I. Ritus, ``The Lagrangian Function of an Intense Electromagnetic
Field'', in {\it Proc. Lebedev Phys. Inst.} Vol. {\bf 168}, {\it Issues
in Intense-field Quantum Electrodynamics}, V. I. Ginzburg, ed., (Nova
Science Pub., NY 1987).



\bibitem{ditreu-book}
W. Dittrich and M. Reuter, {\it Effective Lagrangians in Quantum Electrodynamics}, Springer 1985.


\bibitem{rss}
M. Reuter, M. G. Schmidt and C. Schubert,
Ann. Phys. (N.Y.)
{\bf 259} (1997) 313, hep-th/9610191.

\bibitem{frss}
D. Fliegner, M. Reuter, M. G. Schmidt, C. Schubert,
Teor. Mat. Fiz. {\bf 113} (1997) 289
[Theor. Math. Phys. {\bf 113} (1997) 1442], hep-th/9704194.

\bibitem{korsch}
B. K{\"o}rs, M.G. Schmidt,
Eur. Phys. J. {\bf C 6} (1999) 175,
hep-th/9803144.


\bibitem{dufish} M.J. Duff and C.J. Isham,
Phys. Lett. {\bf 86B}
(1979) 157;
Nucl. Phys. {\bf B 162} (1980) 271.

\bibitem{bardeen}
W.A. Bardeen, Preprint FERMILAB-CONF-95-379-T.

\bibitem{selivanov}
A.A. Rosly, K.G. Selivanov,
Phys. Lett. {\bf B 399} (1997) 135, hep-th/9611101.

\bibitem{cangemi}
D. Cangemi, Nucl. Phys. {\bf B484} (1997) 521,
hep-th/9605208;
Int. J. Mod. Phys. {\bf A 12} (1997) 1215, hep-th/9610021.

\bibitem{chasie}
G. Chalmers, W. Siegel,
Phys. Rev. {\bf D 54} (1996) 7628, hep-th/9606061.

\bibitem{50} 
G.V. Dunne and C. Schubert, 
%``Closed-form two-loop Euler-Heisenberg Lagrangian in a self-dual background'', 
Phys. Lett. {\bf B526} 55 (2002) hep-th/0111134.

\bibitem{51}
G. V. Dunne and C. Schubert, 
%ï¿œTwo-loop self-dual Euler-Heisenberg Lagrangians. I: Real part and helicity amplitudes,ï¿œ 
JHEP {\bf 0208} 053 (2002), hep-th/0205004. 


\bibitem{frgish-book}
E.S. Fradkin, D.M. Gitman, S.M. Shvartsman, {\it Quantum Electrodynamics with Unstable Vacuum}, Springer 1991.

\bibitem{giekar}
H. Gies and F. Karbstein,
%``An Addendum to the Heisenberg-Euler effective action beyond one loop'',
{\it JHEP} {\bf 1703} (2017) 108;  arXiv:1612.07251 [hep-th]. 

\bibitem{112}
J. P. Edwards and C. Schubert,
%``One-particle reducible contribution to the one-loop scalar propagator in a constant field'',
Nucl. Phys. B {\bf 923} (2017) 339, arXiv: 1704.00482 [hep-th].

\bibitem{113}
N, Ahmadiniaz, F. Bastianelli, O. Corradini, J. P. Edwards and C. Schubert,
%``One-particle reducible contribution to the one-loop spinor propagator in a constant field'',
Nucl. Phys. B {\bf 924} (2017) 377, arXiv: 1704.05040 [hep-th].

\bibitem{karbstein-vp}
F. Karbstein,
 %``Tadpole diagrams in constant electromagnetic fields'',
JHEP {\bf 1710} (2017) 075, arXiv:1709.03819 [hep-th].

\bibitem{thooft} G. 't Hooft, ``Computation of the quantum effects due to
a four-dimensional pseudoparticle'', Phys. Rev. {\bf D 14} (1976) 3432.


\bibitem{dadda} A. D'Adda and P. Di Vecchia, ``Supersymmetry and
instantons'', Phys. Lett. {\bf 73B} (1978) 162.

\bibitem{brolee} L. S. Brown and C. Lee, ``Massive propagators in instanton
fields'', Phys. Rev. D {\bf 18} (1978) 2180.




% Please avoid comments such as "For a review'', "For some examples",
% "and references therein" or move them in the text. In general,
% please leave only references in the bibliography and move all
% accessory text in footnotes.

% Also, please have only one work for each \bibitem.


\end{thebibliography}
\end{document}